\documentclass{emulateapj}
\usepackage[]{natbib}
\usepackage{graphics}
\usepackage{apjfonts}
     
\newcommand{\etal}{et al.}  
\newcommand{\per}{\ensuremath{^{-1}}}
\newcommand{\persq}{\ensuremath{^{-2}}}

\newcommand{\percucm}{cm\ensuremath{^{-3}}}
\newcommand{\hal}{H\ensuremath{\alpha}}
\newcommand{\hbeta}{H\ensuremath{\beta}} 

\newcommand{\msun}{\ensuremath{M_{\odot}}}
 
\newcommand{\kms}{km s\ensuremath{^{-1}}}

\newcommand{\mbh}{\ensuremath{M_\mathrm{BH}}}

\newcommand{\chisq}{\ensuremath{\chi^2}}

\newcommand{\sigmastar}{\ensuremath{\sigma_\star}}

\newcommand{\msigma}{\ensuremath{\mbh-\sigmastar}}
\newcommand{\rblr}{\ensuremath{r_{\mathrm{BLR}}}}
\newcommand{\vfwhm}{\ensuremath{v_{\mathrm{FWHM}}}}
\newcommand{\mgb}{Mg~\ensuremath{b}}
\newcommand{\rosat}{\emph{ROSAT}}
\newcommand{\lbol}{\ensuremath{L_{\mathrm{bol}}}}
\newcommand{\lbul}{\ensuremath{L_{\mathrm{bul}}}}
\newcommand{\ledd}{\ensuremath{L_{\mathrm{Edd}}}}

\slugcomment{To Appear in The Astrophysical Journal}
\shorttitle{POX 52} 
\shortauthors{BARTH ET AL.}

\begin{document} 

\title{POX 52: A Dwarf Seyfert 1 Galaxy With An Intermediate-Mass
Black Hole}

\author{Aaron J. Barth\altaffilmark{1,2}, 
Luis C. Ho\altaffilmark{3}, Robert E. Rutledge\altaffilmark{4,5},
and Wallace L. W. Sargent\altaffilmark{1}}

\altaffiltext{1}{Department of Astronomy, 105-24 Caltech, Pasadena, CA
91125}

\altaffiltext{2}{Hubble Fellow}

\altaffiltext{3}{The Observatories of the Carnegie Institution of
  Washington, 813 Santa Barbara Street, Pasadena, CA 91101}

\altaffiltext{4}{Theoretical Astrophysics, 130-33 Caltech, Pasadena,
  CA 91125}

\altaffiltext{5}{Physics Department, McGill University}

\begin{abstract}

We describe new optical images and spectra of POX 52, a dwarf galaxy
with an active nucleus that was originally detected in the POX
objective-prism survey.  While POX 52 was originally thought to be a
Seyfert 2 galaxy, the new data reveal an emission-line spectrum very
similar to that of the dwarf Seyfert 1 galaxy NGC 4395, with broad
components to the permitted line profiles, and we classify POX 52 as a
Seyfert 1 galaxy.  The host galaxy appears to be a dwarf elliptical,
and its brightness profile is best fit by a S\'ersic model with an
index of $3.6\pm0.2$ and a total magnitude of $M_V = -17.6$.  Applying
mass-luminosity-linewidth scaling relations to estimate the black hole
mass from the broad H$\beta$ linewidth and nonstellar continuum
luminosity, we find $M_{\mathrm{BH}} \approx 1.6\times10^5 ~M_{\sun}$.
The stellar velocity dispersion in the host galaxy, measured from the
\ion{Ca}{2} $\lambda8498, 8542$ \AA\ lines, is $36\pm5$ km s$^{-1}$,
also suggestive of a black hole mass of order $10^5 ~M_{\sun}$.
Further searches for active nuclei in dwarf galaxies can provide
unique constraints on the demographics of black holes in the mass
range below $10^6 ~M_{\sun}$.
\end{abstract}


\section{Introduction}

\begin{figure*}
\begin{center}
\rotatebox{-90}{\scalebox{0.66}{\includegraphics{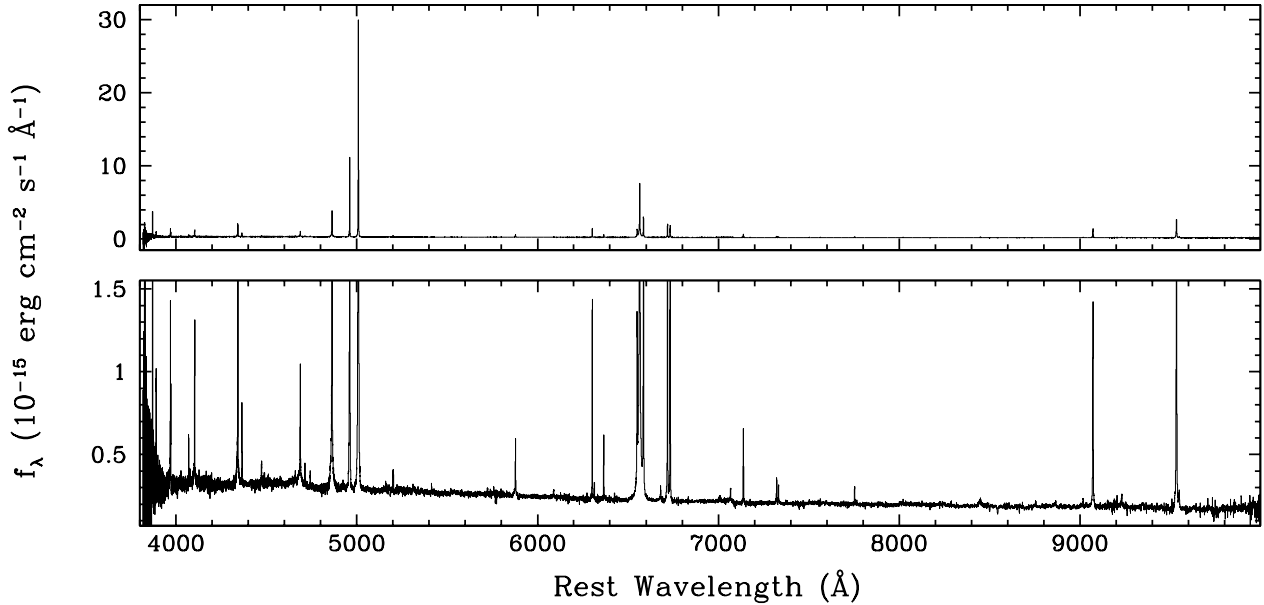}}}
\end{center}
\caption{Keck ESI spectrum of POX 52. 
\label{fullspec}}
\end{figure*}

Studies of the dynamics of stars and gas in the nuclei of nearby
galaxies have detected the signature of supermassive black holes in an
ever-increasing number of galaxies \citep[for recent reviews,
see][]{kor04, bar04}, and it is now widely accepted that all galaxies
with a massive bulge component contain a central black hole.  While
existing surveys have made great progress in the black hole census for
masses in the range $\sim2\times10^6 - 3\times10^9$ \msun, very little
is known about the population of black holes with masses below $10^6$
\msun.  Stellar-dynamical and gas-dynamical searches do not have
sufficient sensitivity to detect black holes of mass $\lesssim10^6$
\msun\ for galaxies much beyond the Local Group.
``Intermediate-mass'' black holes with masses of $\sim10^3 - 10^6$
\msun\ might be present in the centers of very late-type spiral
galaxies, in dwarf elliptical and dwarf spheroidal galaxies, and
possibly even in massive globular clusters, but thus far dynamical
studies have only been carried out for a few very nearby objects,
sometimes with ambiguous results \citep{vdm04}.  At present, there is
little hope of obtaining a systematic census of black holes with $M <
10^6$ \msun\ by using traditional dynamical detection techniques.  The
demographics of intermediate-mass black holes are of particular
importance for future gravitational-wave studies, since it is expected
that the most common massive black hole merger events detected by the
space-based \emph{LISA} interferometer will be in the $10^5-10^6$
\msun\ range \citep{hug01}.

Intermediate-mass black holes that would be undetectable by dynamical
measurements might still reveal their presence by their accretion
luminosity.  The detection of a low-luminosity active galactic nucleus
(AGN) in a dwarf galaxy would be a good indication that a black hole
is present, and indirect methods could be used to estimate the black
hole mass even if a stellar-dynamical mass measurement is impossible.
However, only two AGNs in dwarf galaxies have been identified
previously.  The better-known case is the nearby ($D \approx 4$ Mpc),
late-type, dwarf spiral galaxy NGC 4395.  \citet{fs89} found that NGC
4395 contains the least luminous known Seyfert 1 nucleus.  Despite the
fact that this galaxy has no discernible bulge, and thus might not be
expected to contain a central black hole, its nucleus exhibits
all of the characteristics of Seyfert activity, including broad
emission lines \citep{fs89}, a compact, unresolved nonstellar
continuum source in the optical and ultraviolet \citep{fhs93},
unresolved radio emission with a high brightness temperature
\citep{wfh01}, and rapid X-ray variability \citep{iwa00, sif03,
mor04}.  The central stellar velocity dispersion is $\sigmastar < 30$
\kms, which combined with the radius of the central star cluster (3.9
pc) gives a firm upper limit of $M < 6\times10^6$ \msun\ for the
combined mass of the central star cluster plus black hole
\citep{fh03}.  The black hole itself has a likely mass of
$\sim10^4-10^5$ \msun, based on the X-ray variability timescale
\citep{sif03} and extrapolation of the linewidth-luminosity-mass
correlations for Seyfert galaxies \citep{fh03}.  Finding more active
galaxies like NGC 4395 would be a valuable step toward understanding
the demographics of black holes with $M < 10^6$ \msun.

POX 52 (also known as PGC 038055 or G1200--2038) is a dwarf galaxy at
$cz = 6533$ \kms.  It was discovered in the course of an
objective-prism search for emission-line objects performed by
\citet{ksk81}.  They listed the fluxes for several emission lines in
POX 52 and noted that it appears ``star-like'' in Palomar Sky Survey
images.  Follow-up spectroscopy and imaging data were presented by
\citet[][hereafter KSB]{ksb87}.  KSB called attention to POX 52 as an
unusual example of a dwarf galaxy having an AGN spectrum.  The object
was classified as a Seyfert 2 based on the narrow-line ratios,
although they noted that the \hbeta\ line had a weak broad component
with full-width at half-maximum (FWHM) $\approx840$ \kms.  They also
found a weak broad component to the \ion{He}{2} $\lambda4686$ emission
line, which they tentatively ascribed to emission from Wolf-Rayet
stars. The host galaxy was found to have an exponential scale length
of only 1 kpc and an absolute magnitude of $M_V=-16.9$.  This is a
surprisingly small host galaxy for an AGN.  However, a literature
search reveals that no additional optical observations of POX 52 have
been published since the initial study by KSB.  Motivated by the
possibility that POX 52 might contain an intermediate-mass black hole
similar to the one in NGC 4395, we obtained new spectra and images of
POX 52 at the Keck and Las Campanas observatories.\footnote{At the
time of this writing, the SIMBAD database contains incorrect
coordinates for POX 52.  The correct J2000 coordinates are listed in
NED as $\alpha=12^\mathrm{h}02^\mathrm{m}56\fs9$, $\delta=
-20\arcdeg56\arcmin03\arcsec$.}

Throughout this paper we assume a Hubble constant of $H_0 = 70$ km
s\per\ Mpc\per.

\section{Observations and Reductions}

\subsection{Spectroscopy}

We observed POX 52 with the ESI spectrograph \citep{she02} at the
Keck-II telescope on the night of 2002 December 2 UT.  In echellette
mode, ESI gives complete coverage of the wavelength range 3800--11000
\AA\ in 10 echelle orders, at a pixel scale of 11.5 km s\per\ pix\per.
We used a 0\farcs5-wide slit, for a FWHM resolution of $\sim36$ km
s\per\ as measured from the widths of arc lamp lines.  The total
exposure time for POX 52 was 2400 s, and the airmass ranged from 1.85
to 1.65 during the observations.  The slit was oriented at PA =
320\arcdeg, which was the parallactic angle for the midpoint of the
exposure.  This angle coincided almost exactly with the position angle
of the host galaxy major axis, as we later determined from imaging
data.  On the same night, we also observed the nucleus of NGC 4395
with the same setup for 2400 s, at airmass 1.6, as well as 7 giant
stars of spectral type ranging from G9III to K5III for use as velocity
dispersion templates.  The seeing, in standard star exposures taken at
airmass 1.2 immediately after the galaxy observations, was
0\farcs9--1\farcs0.  Wavelength calibration was performed using
observations of HgNe, Xe, and CuAr lamps.

The spectra were flattened, extracted, and wavelength-calibrated using
both the automated MAKEE software
package\footnote{http://www2.keck.hawaii.edu/inst/common/makeewww/index.html}
and also using standard tasks in IRAF\footnote{IRAF is distributed by
the National Optical Astronomy Observatories, which are operated by
the Association of Universities for Research in Astronomy, Inc., under
cooperative agreement with the National Science Foundation.}, as a
comparison.  The extraction width was 2\farcs5. The extractions
performed using the two software packages were virtually identical
except for wavelengths longer than $\sim8000$ \AA, where the sky
subtraction was more accurate in the IRAF reductions.  The extractions
were then flux-calibrated and corrected for telluric absorption bands
using observations of the standard star Feige~34.

\subsection{Imaging}

Optical images of POX 52 were obtained at the du Pont 2.5-m telescope
at Las Campanas Observatory on the night of 2003 February 10 UT, using
the Direct Tek\#5 CCD Camera with a pixel scale of 0\farcs259.  The
night was photometric with $\sim0\farcs75$ seeing.  POX 52 was
observed using \emph{B}, \emph{V}, \emph{R}, and \emph{I} filters,
with an exposure time of $2\times360$ s in each filter.  Two
\citet{lan92} standard star fields were observed for photometric
calibration.  The images were bias-subtracted and flattened, and the
photometric zeropoints were determined using standard tasks in IRAF.

\section{The Active Nucleus}

\subsection{Emission Lines}

Figure \ref{fullspec} displays the complete ESI spectrum of POX 52.
The data confirm KSB's classification of POX 52 as a Seyfert galaxy,
as demonstrated by the large [\ion{O}{3}]/\hbeta\ ratio and
the strengths of [\ion{O}{1}] $\lambda6300$, [\ion{N}{2}]
$\lambda6584$, and [\ion{S}{2}] $\lambda\lambda$6716, 6731 relative to
\hal.    

More than 40 emission lines are visible in the spectrum.  To measure
the line intensities, we first fitted a low-order spline to the
continuum in each echelle order and subtracted it from the spectrum.
The line fluxes were measured by direct integration, or in the case of
some blended lines, by performing a multi-Gaussian model fit, and the
results are listed in Table \ref{emissionlines}.  Measurement
uncertainties are dominated by systematics including the exact
placement of the underlying continuum; we estimate that the fluxes of
the strong lines are uncertain by $\sim5-10\%$ while the uncertainties
on the weakest lines may be $\sim40-50\%$.  In addition, there is an
overall uncertainty to the flux scale because of slit losses, but this
should not affect line ratios.  The Galactic extinction toward POX 52
appears to be slightly uncertain; \citet{bh82} give $A_B = 0.12$ mag
while \citet{sfd98} find $A_B = 0.225$ mag along this line of sight.
As a compromise, we chose to to compute extinction-corrected line
intensities using the mean of these two extinction values, and using
the reddening curve of \citet{ccm98}.  The peak wavelengths of the 12
strongest emission lines give a heliocentric recession velocity of $cz
= 6533 \pm 6$ \kms, while the stellar \mgb\ and \ion{Ca}{2} triplet
absorption lines yield $cz = 6531\pm12$ \kms.  These results differ
substantially from the value of 6420 \kms\ reported by KSB.

\begin{deluxetable}{lccc}
\tablewidth{5in} 
\tablecaption{Emission Lines in POX 52} 
\tablehead{
\colhead{Line} & \colhead{Rest Wavelength} & \colhead{Flux} &
\colhead{$I/I$(\hbeta$_{\mathrm{n}}$)} \\ \colhead{} & \colhead{(\AA)} &
\colhead{($10^{-15}$ erg cm\persq\ s\per)} & \colhead{} }
\startdata 
{}[Ne III]          &   3869.1       &    4.8   &      0.66     \\
He I + H$\zeta$   &3888.6, 3889.0  &     1.0  &      0.14     \\
{}{}[Ne III]        &   3967.8       &     1.7  &      0.23     \\
H$\epsilon$              &   3970.1       &     0.9  &      0.12     \\
{}[S II]           &   4068.6       &     0.5  &     0.07      \\
{}[S II]           &   4076.4       &     0.1  &     0.01      \\
H$\delta$                &   4101.7       &     2.1  &      0.29     \\
H$\gamma$                &   4340.5       &     4.6  &      0.63     \\
{}[O III]           &   4363.2       &     0.9  &      0.12     \\
He I              &   4471.5       &     0.3  &     0.04      \\
{}[Fe III]          &   4658.1       &     0.2  &     0.03      \\
He II              &   4685.7       &     3.0  &      0.40     \\
{}[Ar IV]          &   4711.4       &     0.2  &     0.03      \\
{}[Ar IV]          &   4740.2       &     0.2  &     0.03      \\
\hbeta\ narrow           &   4861.3       &     7.5  &       1.00    \\
\hbeta\ broad            &   4861.3       &     3.9  &      0.52     \\
{}[O III]           &   4958.9       &    20.9  &       2.78    \\
{}[O III]           &   5006.8       &    60.4  &       8.01    \\
{}[Fe VII]          &   5158.4       &     0.1  &     0.01      \\
{}[N I]           &   5197.9       &     0.2  &     0.03      \\
{}[N I]           &   5200.3       &     0.2  &     0.03      \\
He II               &   5411.5       &     0.1 &     0.01      \\
{}[Fe VII]          &   5720.7       &     0.1  &     0.01      \\
{}[N II]           &   5754.6       &     0.1  &     0.01      \\
He I              &   5875.7       &     1.3  &      0.17     \\
{}[Fe VII]          &   6086.3       &     0.2  &     0.03      \\
{}[O I]           &   6300.3       &    2.6   &      0.33     \\
{}[S III]           &   6312.1       &     0.2  &     0.03      \\
{}[O I]           &   6363.8       &     0.9  &      0.12     \\
{}[N II]           &   6548.0       &     2.3  &      0.29     \\
\hal\ narrow             &   6562.8       &    23.4  &       2.98    \\
\hal\ broad              &   6562.8       &    11.1  &       1.41    \\
{}[N II]           &   6583.5       &     7.2  &      0.92     \\
He I              &   6678.2       &     0.4  &     0.05      \\
{}[S II]           &   6716.4       &     5.3  &      0.67     \\
{}[S II]           &   6730.8       &     4.7  &      0.60     \\
{}[Ar V]          &   7005.4       &     0.1  &     0.01      \\
He I              &   7065.3       &     0.3  &     0.04      \\
{}[Ar III]          &   7135.8       &     1.3  &      0.16     \\
{}[O II]           &7318.9, 7320.0  &     0.5  &     0.06      \\
{}[O II]           &7329.7, 7330.7  &     0.4  &     0.05      \\
{}[Ar III]          &   7751.1       &     0.3  &     0.04      \\
O I               &   8446.4       &     0.6  &     0.07      \\
P12                      &   8750.5       &     0.3  &     0.04      \\
P11                      &   8862.8       &     0.2  &     0.02      \\
P10                      &   9014.9       &     0.2  &     0.02      \\
{}[S III]           &   9068.6       &     4.0  &      0.49     \\
P9                       &   9229.0       &     0.3  &     0.04      \\
{}[S III]           &   9530.6       &    10.0  &       1.22    \\
P$\epsilon$              &   9546.0       &     0.4  &     0.05      \\
P$\delta$                &  10049.4       &     0.7  &     0.08      \\
\enddata
\tablecomments{Measured line fluxes do not take into account slit
  losses for the observations of POX 52 or of the standard star, so
  the overall flux scale may be inaccurate although relative fluxes
  should be unaffected.  Intensity ratios $I/I$(\hbeta$_\mathrm{n}$)
  are intensities relative to the narrow component of \hbeta,
  corrected for a Galactic extinction of $A_B$ = 0.173 mag, which is
  the mean of the extinction estimates from \citet{bh82} and from
  \citet{sfd98}.   Except for \hal\ and \hbeta, the measured line
  fluxes are total fluxes for the narrow and broad components
  combined.   The air wavelengths for emission lines are taken
  from the Atomic Line List at
  http://www.pa.uky.edu/$\sim$peter/atomic/ .  }
\label{emissionlines}
\end{deluxetable}

One notable feature of the POX 52 spectrum is the presence of
[\ion{Fe}{7}] emission with lines at 5158, 5721 and 6086 \AA.  These
emission lines are also found in the spectrum of NGC 4395
\citep{kra99}, and some narrow-line Seyfert 1 (NLS1) galaxies
\citep{op85}.  The detection of these high-excitation features further
confirms the classification of POX 52 as a genuine Seyfert galaxy.

\begin{figure}
\begin{center}
\rotatebox{-90}{\scalebox{0.33}{\includegraphics{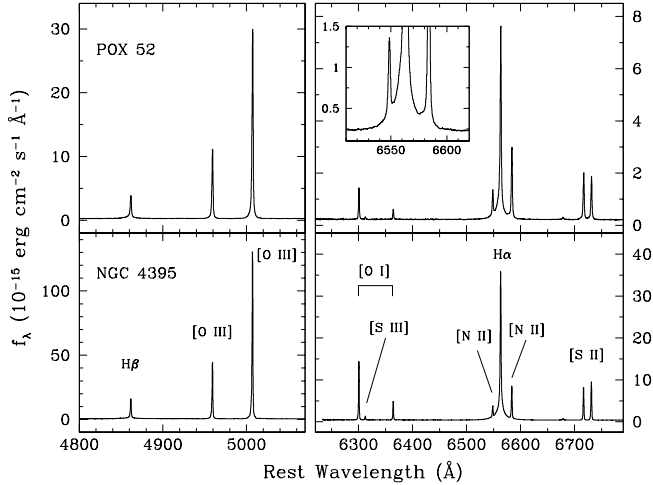}}}
\end{center}
\caption{Keck ESI spectra of POX 52 and NGC 4395.  For each object,
  the two panels show the wavelength ranges containing the \hbeta\ and
  [O III] lines (left panels) and the [O I], \hal,
  [N II], and [S II] lines (right panels).  The inset
  shows a close-up view of the broad base of \hal\ in POX 52.  
\label{specplot}}
\end{figure}

\begin{figure}[t]
\begin{center}
\scalebox{0.4}{\includegraphics{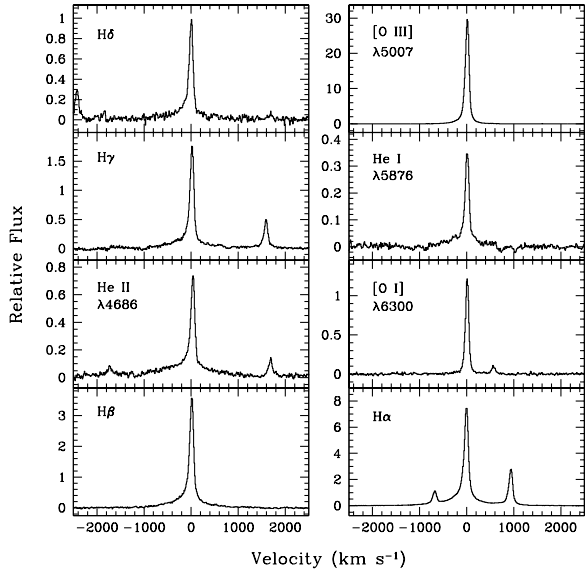}}
\end{center}
\caption{Profiles of emission lines in POX 52.  The spectra have been
  continuum-subtracted using a low-order spline anchored to the
  spectra at velocities beyond $\pm2000$ \kms\ from the emission-line
  centers. Some residual stellar absorption features are visible.
\label{lineprofiles}}
\end{figure}

Figure \ref{specplot} shows the spectral regions around [\ion{O}{3}]
and \hbeta, and the region surrounding \hal, for POX 52 and NGC 4395.
The two objects have extremely similar spectra, with only small
differences in the line ratios and widths.  POX 52 has broad wings on
the \hal\ emission line that appear nearly identical to those in NGC
4395.  KSB noted the presence of a weak broad component to the \hbeta\
and \hal\ lines but still classified POX 52 as a Seyfert 2 galaxy.
The broad components are clearly present in the Keck data; thus, POX
52 should be classified as a Seyfert 1 galaxy.  In the \citet{ost81}
classification scheme, POX 52 would be considered a Seyfert 1.8
galaxy; NGC 4395 falls in the same category \citep{hfs97}.  Figure
\ref{lineprofiles} displays the profiles of several permitted lines as
well as [\ion{O}{3}] $\lambda5007$ and [\ion{O}{1}] $\lambda6300$.
The broad wings are visible on the profiles of H$\gamma$, H$\beta$,
\hal, \ion{He}{2} $\lambda4686$, and \ion{He}{1} $\lambda5876$.  In
contrast, the [\ion{O}{1}] and [\ion{O}{3}] lines appear devoid of
broad components.

\begin{figure}[t]
\begin{center}
\scalebox{0.4}{\includegraphics{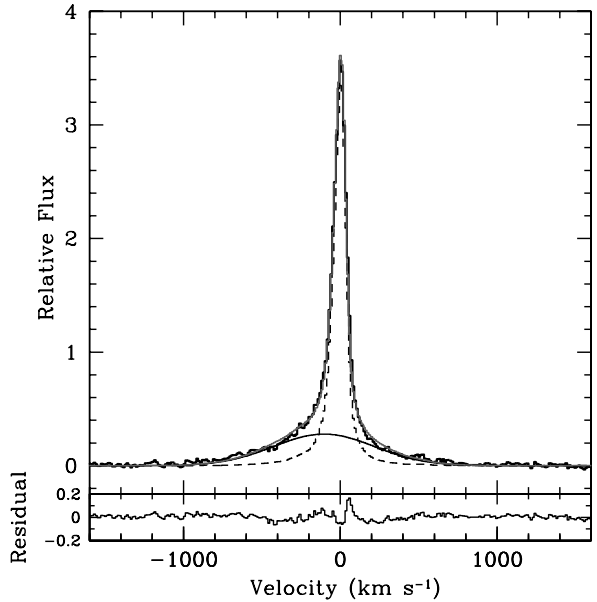}}
\end{center}
\caption{Model fit to the \hbeta\ line, including a scaled version of
  the [O III] $\lambda5007$ profile (dashed line) for the narrow
  component and a Gaussian for the broad component.  The residuals of
  the best-fitting model are shown in the lower panel.
\label{hbetafit}}
\end{figure}

To decompose the \hbeta\ profile into narrow and broad components, we
performed a fit to the continuum-subtracted \hbeta\ line using a model
consisting of a broad Gaussian plus a narrow component with the same
profile as the [\ion{O}{3}] $\lambda5007$ line.  The free parameters
in the fit included the flux and central wavelength of the narrow-line
profile, and the total flux, central wavelength, and FWHM of the broad
Gaussian.  Figure \ref{hbetafit} displays the best-fitting model,
which has a broad-line width of FWHM(\hbeta) = $760 \pm 30$ \kms.  (We
note that this is quite close to the measurement of FWHM = 840 \kms\
by KSB.)  The need for a distinct broad component in the model fit is
clearly apparent, and the broad component contains 34\% of the total
\hbeta\ flux.

We performed a similar decomposition of the \hal\ profile, using a
Gaussian for the broad component and the [\ion{O}{3}] profile to
represent the narrow \hal\ component and the [\ion{N}{2}] lines.  The
results are more uncertain than for the \hbeta\ model fit because of
the contribution from the [\ion{N}{2}] emission lines, and we estimate
that the broad component flux for \hal\ listed in Table
\ref{emissionlines} may be uncertain by $\sim20\%$.  However, the
\emph{total} (broad + narrow) \hal/\hbeta\ intensity ratio (which is
insensitive to the broad/narrow decomposition), corrected for Galactic
extinction, is $I($\hal$)/I($\hbeta$) = 2.9$, indicating that there is
little internal extinction within POX 52.  The width of the broad-line
component is also rather uncertain because the model is a poor fit to
the extended wings of \hal, but the best-fitting model has FWHM = $765
\pm 80$ \kms, consistent with the broad \hbeta\ linewidth.

Overall, the narrow-line intensities relative to \hbeta\ are very
similar to those reported by \citet{kra99} for NGC 4395.  One
difference, which is apparent in Figure \ref{specplot}, is the ratio
of the [\ion{S}{2}] $\lambda\lambda6716, 6731$ emission lines,
indicating a lower-density narrow-line region (NLR) in POX 52.  The
ratio $I(6716)/I(6731)$ is 1.12 in POX 52 and 0.88 in NGC 4395,
corresponding to electron densities of $n_e \approx 360$ and 1000
\percucm, respectively, for $T_e = 10^4$ K.  The lower
[\ion{O}{1}]/[\ion{S}{2}] ratio is also indicative of a lower-density
NLR in POX 52.  The larger [\ion{N}{2}]/\hal\ ratio in POX 52 suggests
that it may have a higher nitrogen abundance than NGC 4395;
\citet{kra99} have shown that the NGC 4395 nucleus has a low N/O
ratio of about 1/3 solar.

KSB found that the widths of the narrow lines in POX 52 were
marginally resolved in comparison with their arc lamp spectrum, and
inferred an intrinsic linewidth of FWHM = 320 \kms\ for the narrow
emission lines in POX 52.  The ESI data show that the forbidden lines
are much narrower than this estimate.  The [\ion{O}{3}] $\lambda5007$
linewidth was measured in three different ways: by a direct
measurement of the full width at half of the peak flux (FWHM = 93
\kms), by a Gaussian fit (FWHM = 102 \kms), and by a Lorentzian fit
(FWHM = 83 \kms).  Neither model fit proved to fit the asymmetric line
profile particularly well, and we adopt FWHM = $93 \pm 10$ \kms\ as
the best estimate of the observed linewidth.  Subtracting the
instrumental FWHM of 0.56 \AA\ (measured from arc lamp exposures) in
quadrature, the intrinsic linewidth is then FWHM([\ion{O}{3}]) = $87
\pm 10$ \kms.  

\subsection{X-ray and Radio Data}

\newcommand{\bob}{\bf}

As a Seyfert 1 galaxy, POX 52 should contain a central X-ray source.
We searched the available X-ray
archives\footnote{http://heasarc.gsfc.nasa.gov} for images containing
the position of POX 52, and found only one matching dataset, from the
\rosat\ all-sky survey \citep{vog99}.  This survey scanned the sky
with the \rosat\ PSPC detector (0.4-2.4 keV energy band), covering
each point on average with a $\sim300$ s integration.  The typical
1$\sigma$ positional uncertainty is 12\arcsec, which includes a
systematic uncertainty of 6\arcsec, largely due to boresight
corrections.  The FWHM of the PSPC point-spread
function (PSF) is 20\arcsec.  The exposure time for the field
surrounding POX 52 was only $90\pm2$ s, where the uncertainty is the
range of exposure in adjacent 15\arcsec\ cells.

\begin{figure}
\begin{center}
\scalebox{0.5}{\includegraphics{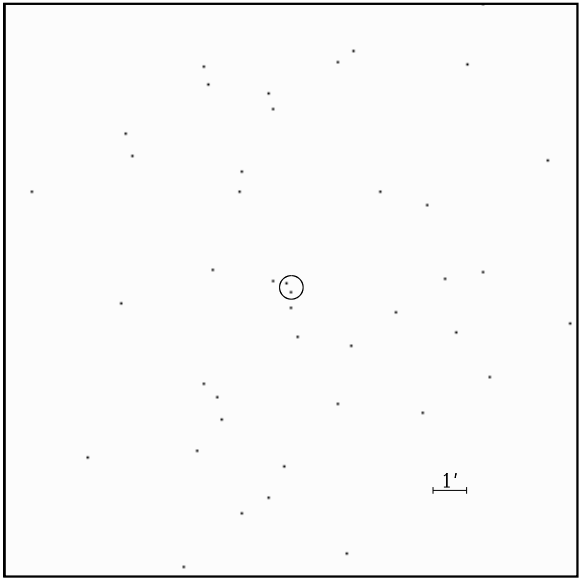}}
\end{center}
\caption{A portion of the \rosat\ All-Sky Survey image of the field
  surrounding POX 52.  Each point corresponds to one X-ray count.  The
  error circle is centered on the position of POX 52 and has $r =
  21\arcsec$.  North is up, and east is to the left.
\label{rosat}}
\end{figure}

There is no obvious, strong X-ray source at the position of POX 52,
but careful examination of the image does reveal a localized excess
above the background count rate.  The 75\% energy radius (that is, the
radius which would enclose 75\% of the PSPC PSF) is 20\arcsec\
\citep{boe00}.
Combining the 20\arcsec\ radius for the PSF with the 6\arcsec\
systematic uncertainty (summed in quadrature), we took a region of
radius 21\arcsec\ about the known position of POX 52 in the \rosat\
data, and found two counts within it (see Figure \ref{rosat}).  In the
surrounding 1800\arcsec\ radius region, there were 370 counts, giving
a mean background level of $\rho = 3.6 \times 10^{-5}$ counts per
square arcsecond.  Then, the probability of two background events
falling within an error circle with $r = 21\arcsec$ is
$P_2(r=21\arcsec)=[1-\exp{(-\rho\pi\, r^2)}]^2 = 0.2\%$.  If we permit
the error circle to expand to $r = 30\arcsec$ (at which point there is
a third count) the probability of chance coincidence of two counts at
this location can be as high as 0.9\%.

Examination of the properties of the two counts showed them to be of
low energy, occurring in pulse-height invariant (PI) channels 57 and
58, at approximately $0.5$ keV.  Taking only PI channels 20-80 for
background, this decreases the background rate to $1.7\times10^{-5}$
counts per square arcsecond.  Subsequently, the significance for two
counts within 21\arcsec\ becomes 0.04\%, and within 30\arcsec\ is
0.2\%.  We adopt the significance of detection to be in the range of
0.04\%-0.2\% (3.0-3.5$\sigma$).

The observed energy fluence from this source is approximately 1
keV. Using Poisson statistics to determine the counting uncertainty
\citep{geh86}, the detection contains 2$^{+2.6}_{-1.9}$ counts
($1\sigma$ Poisson error bars).  The detector area at 0.5 keV is
$\sim500$ cm$^{2}$, and total flux from the source is
$(4.0^{+5.2}_{-3.8})\times10^{-14}$ erg cm\persq\ s\per.  For $D = 93$
Mpc, this corresponds to a time-averaged luminosity of
$(4.3^{+5.6}_{-4.1})\times10^{40}$ erg s\per.  Note that the detection
certainty is higher than indicated by the flux uncertainty.

Unfortunately, given the shallowness of the \rosat\ exposure and the
bare 3.0-3.5$\sigma$ significance of the source detection, the best
that can be done is to extract a very rough estimate of the X-ray
luminosity.  It would be extremely worthwhile to observe POX 52 with
\emph{Chandra} or \emph{XMM-Newton} to confirm this detection, and to
properly quantify its X-ray luminosity, spectral shape, and
variability, particularly in light of the extreme short-timescale
variability recently found in NGC 4395 \citep{sif03, mor04}.

A \rosat\ observation of NGC 4395 by \citet{mor99} revealed a soft
X-ray luminosity of $3.4\times10^{38}$ erg s\per\ (after rescaling to
$D = 4.2$ Mpc).  Taking our \rosat\ measurement at face value, POX 52
is then $\sim100$ times more luminous than NGC 4395 in soft X-rays,
although this is at best a rough comparison because of the large
uncertainty in the \rosat\ measurement as well as the possibility of
strong X-ray variability in POX 52.

We searched for POX 52 in the NRAO VLA Sky Survey archives
\citep{con98} and did not find a source at the galaxy's position.
This implies a $3\sigma$ upper limit of $\sim1.4$ mJy at 1.4 GHz.  The
field of POX 52 was not observed in the VLA FIRST survey.

\section{The Host Galaxy}

\subsection{Stellar Velocity Dispersion}

Numerous stellar absorption features are visible in the ESI spectrum,
making it possible to measure the stellar velocity dispersion
\sigmastar.  We determined \sigmastar\ from the \ion{Ca}{2}
near-infrared triplet lines, which are generally the best features for
measurement of stellar kinematics \citep{dre84}, and also from the
region surrounding \mgb.

\begin{figure}
\begin{center}
\scalebox{0.32}{\rotatebox{-90}{\includegraphics{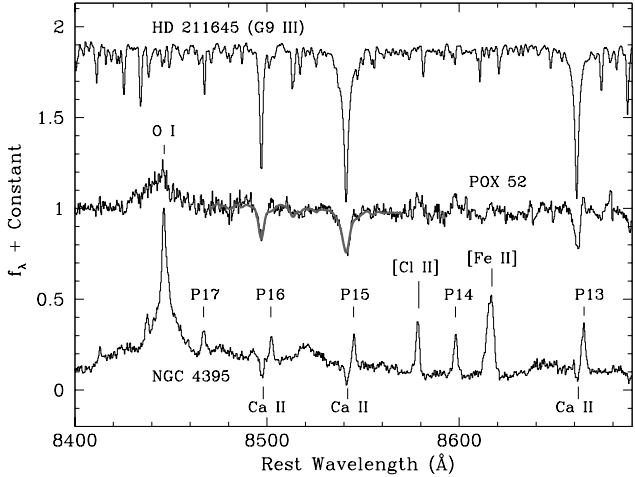}}}
\end{center}
\caption{The Ca II triplet spectral region in POX 52, NGC 4395, and
  the G9 III star HD 211645.  All spectra have been rescaled to a
  median flux level of unity, and the spectra of HD 211645 and NGC
  4395 were then shifted vertically for clarity.  The smooth curve
  overplotted on the spectrum of POX 52 is the stellar spectrum after
  broadening and dilution by the featureless continuum.  The
  identification of [Cl II] $\lambda8578.7$ in NGC 4395 is supported
  by the additional detection of [Cl II] $\lambda9123.6$ in the same
  spectrum.  [Cl~II] $\lambda8578.7$ emission appears to be very
  weakly present in POX 52 as well.  This line has previously been
  detected in NGC 4151 \citep{osv90} and NGC 1068 \citep{of96}.
\label{dispersion}}
\end{figure}

\citet{fh03} have recently published high-resolution spectra of the Ca
triplet spectral region for NGC 4395 and determined an upper limit to
its stellar velocity dispersion ($\sigmastar < 30$ \kms).  The
spectrum of NGC 4395 presented particular problems for this
measurement, because the \ion{Ca}{2} $\lambda8542$ and $\lambda8662$
absorption features were almost completely filled in by the
Paschen-series P16 and P15 emission lines.  This left \ion{Ca}{2}
$\lambda8498$ as the only uncontaminated stellar feature that could be
used to measure \sigmastar.  The P16 emission line is located 6 \AA\
redward of \ion{Ca}{2} $\lambda8498$, but in NGC 4395 the emission and
absorption lines are sufficiently narrow that the two features do not
overlap.  The presence of higher-order Paschen emission lines is an
unusual characteristic of NGC 4395 and POX 52, which have extremely
narrow emission lines of relatively high equivalent width.  In most
Seyfert galaxies the high-order P13--P17 lines are not visible at all;
it is much more common to see the \ion{Ca}{2} lines themselves in
emission, particularly in NLS1 galaxies having strong \ion{Fe}{2}
emission \citep[e.g.,][]{per88}.

In POX 52, some high-order Paschen-series emission lines are visible (Table
\ref{emissionlines}) but their equivalent widths are smaller than
those in NGC 4395, as shown in Figure \ref{dispersion}.  The
\ion{Ca}{2} $\lambda8662$ line is partly filled in by P13 emission,
but the 8498 and 8542 \AA\ absorption lines appear to be largely
uncontaminated, and we use those lines to determine \sigmastar.

The measurement was performed by a direct fit of Gaussian-broadened
stellar spectra to the POX 52 spectrum over the range 8470--8570 \AA,
following the techniques described by \citet{bhs02}.  The
model-fitting code broadens the stellar template by convolution with a
Gaussian in velocity space, and allows an additive featureless
continuum contribution and a multiplicative correction described by a
low-order polynomial (a quadratic in this case) to account for
reddening or other differences in overall spectral shape between the
galaxy and the template star.  The fit of the broadened template star
to the galaxy spectrum is performed directly in pixel space so that
individual spectral features such as emission lines can be masked out
from the computation of \chisq.  The result for POX 52 is $\sigmastar
= 36 \pm 5$ \kms, where the uncertainty is the sum in quadrature of
the fitting error for the best-fitting template star (HD 211645, G9
III) and the standard deviation of the velocity dispersions measured
from all 7 template stars.  The model fits include a featureless
continuum that contributes $\sim60-65\%$ of the total flux in the
extracted spectrum at 8500 \AA\ to match the dilution of the stellar
features.  One concern is that weak Paschen emission could be partly
filling in the red wings of the \ion{Ca}{2} 8498 and 8542 \AA\
absorption lines, which would make the above result an underestimate
of the true velocity dispersion.  To test this possibility, we
performed the measurement again, fitting only to the core and blue
wing of each absorption line, and the surrounding continuum, but
excluding the red wing of both Ca lines from the fit (i.e., the
regions that would be most strongly affected by Paschen emission).
The best-fitting template star then gave $\sigmastar = 39\pm7$ \kms,
in agreement with the $1\sigma$ uncertainty range of our original
measurement.  Thus, the result $\sigmastar = 36\pm5$ \kms\ appears to
be robust even if there is slight contamination by the P15 and P16
emission lines.

\begin{figure}
\begin{center}
\scalebox{0.32}{\rotatebox{-90}{\includegraphics{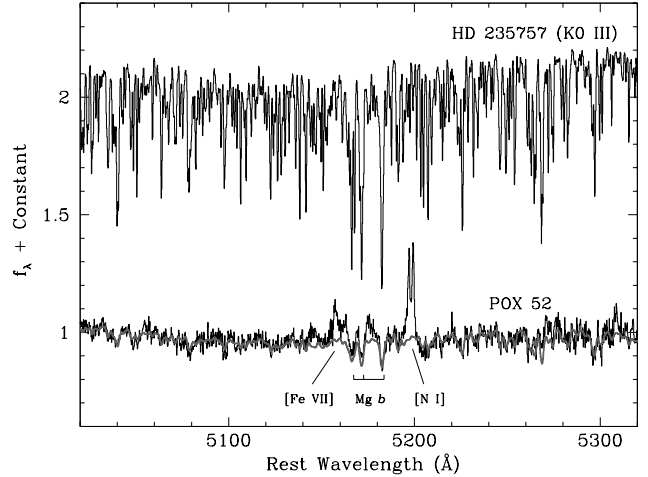}}}
\end{center}
\caption{The spectral region surrounding the \mgb\ lines in POX 52 and
  in the K0 III star HD 235757.  Both spectra were rescaled to a
  median flux level of unity, and the spectrum of HD 235757 was then
  shifted vertically by 1.0 units for clarity.  The smooth curve
  overplotted on the spectrum of POX 52 is the stellar template after
  broadening and dilution by the featureless continuum.  The region
  5140--5210 \AA\ was excluded from the calculation of \chisq\ in the
  model fitting, because of the presence of emission lines.  
\label{mgb}}
\end{figure}

As an additional check on this result, we also performed model fits to
the spectral region surrounding the \mgb\ absorption lines.  As shown
in Figure \ref{mgb}, the \mgb\ lines are contaminated by emission from
[\ion{Fe}{7}] $\lambda5158$ and from [\ion{N}{1}] $\lambda\lambda5198,
5200$, as well as a broad ``bump'' in the spectrum over
$\sim5160-5180$ \AA\ which may be a blend of weaker emission lines.
Therefore, we fit broadened stellar templates to the regions
immediately surrounding \mgb, from 5020 to 5320 \AA\, but excluding
the region 5140--5210 \AA\ from the \chisq\ minimization.  The result,
$\sigmastar = 38 \pm 4$ \kms, is consistent with the dispersion
measured from the \ion{Ca}{2} lines, and the broadened template gives
a reasonable fit to the galaxy spectrum except at wavelengths affected
by the emission lines.

\subsection{Photometric Decomposition}

\begin{figure}
\begin{center}
\scalebox{0.5}{\includegraphics{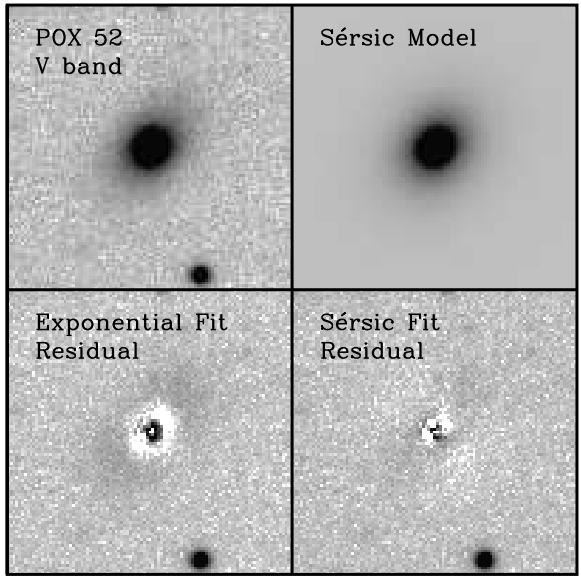}}
\end{center}
\caption{$V$-band image of POX 52.  The image size is 25\arcsec\ on a
  side.  The four panels show the original data, the best-fitting
  S\'ersic model for the host galaxy, the residual from the best
  fitting exponential + point source model, and the residual from the
  best-fitting S\'ersic + point source model.  Each image is displayed
  with the same logarithmic color stretch.  The images are oriented so
  that north is up, and east is to the left.
\label{galfitfig}}
\end{figure}

KSB found that the luminosity profile of POX 52 was best represented
by an exponential disk model with a central point source, and we
expected that our optical imaging data would show spiral structure or
evidence for recent star formation in a host galaxy similar to NGC
4395.  However, the new images reveal a very different morphology for
POX 52.  There is no obvious spiral or disklike structure in the
images, and none of the images shows any obvious clumpiness such as
might be expected from star-forming regions in the host galaxy.  As
shown in Figure \ref{galfitfig}, the optical structure of POX 52 is
most suggestive of a small elliptical galaxy with a bright nuclear
point source.

To study the structure of the host galaxy quantitatively, we fit
models to the galaxy's luminosity profile using the 2-dimensional
fitting code GALFIT \citep{peng02}.  GALFIT is designed to optimize
model fits to galaxy images, allowing multicomponent models with
bulge, disk, and point-source components, and including convolution of
the model by the telescope point-spread function.  The PSF for
each filter was determined from about 10 unsaturated stars in each
image, using standard routines in IRAF.  However, the nucleus of POX
52 was brighter than any of the unsaturated stars in the image, and
the resulting PSF models proved to be somewhat noisy.  We created
noise-free model PSFs by fitting a 2-dimensional Moffat function to
the empirical PSF for each filter, and the model PSFs were then used
in the GALFIT fitting.

The fitting was performed for each filter using two different galaxy
models: an exponential profile plus central point source, and a
S\'ersic $r^{1/n}$ profile \citep{ser68} with index $n$ allowed to
vary, plus central point source.  For each model fit, the free
parameters included the host galaxy scale length or effective radius
$r_e$, magnitude, centroid, major axis position angle, and
ellipticity, and the point source magnitude and centroid.  A
boxiness/diskiness parameter for the host galaxy was also allowed to
vary.  Figure \ref{galfitfig} shows the fitting results for the
\emph{V} band; we found similar results for the other bands.  In each
case, we found that the exponential profile was an extremely poor
model for the host galaxy structure, and the best-fitting exponential
profile plus point source model left a characteristic residual pattern
of concentric positive and negative ``rings''.

\begin{deluxetable}{lcc|cccc}[t]
\tablewidth{4.5in} 
\tablecaption{Results from Galaxy Profile Fits} 

\tablehead{ \colhead{Band} & \multicolumn{2}{c}{Point Source} &
  \multicolumn{4}{c}{Host Galaxy} \\
\colhead{} & \colhead{$m_\mathrm{point}$} &
  \colhead{$M_\mathrm{point}$} & \colhead{$m_\mathrm{host}$} &
  \colhead{$M_\mathrm{host}$} & \colhead{S\'ersic $n$} &
  \colhead{$r_e$ (\arcsec)}
} 

\startdata

\emph{B} & 18.3 & $-16.7$ & 18.3 & $-16.8$ & 3.9 & 1.2 \\ 
\emph{V} & 17.9 & $-17.1$ & 17.4 & $-17.6$ & 3.5 & 1.2 \\
\emph{R} & 17.6 & $-17.4$ & 16.7 & $-18.2$ & 3.7 & 1.2 \\
\emph{I} & 17.4 & $-17.5$ & 16.3 & $-18.6$ & 3.4 & 1.5 \\

\enddata \tablecomments{Apparent magnitudes ($m$) are not corrected
for extinction.  Absolute magnitudes ($M$) are corrected for
extinction of $A_B = 0.173$ mag and computed for $D = 93$ Mpc.}
\label{galfitresults}
\end{deluxetable}

\begin{figure}
\begin{center}
\scalebox{0.4}{\includegraphics{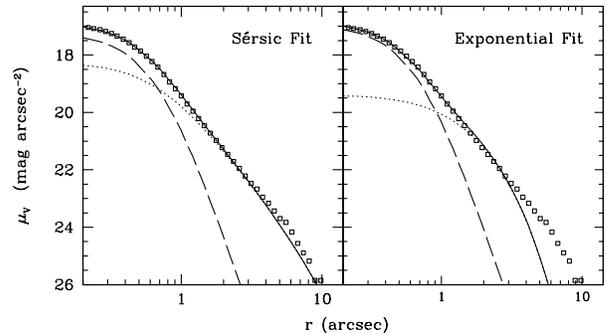}}
\end{center}
\caption{$V$-band Radial profile of POX 52, compared with a S\'ersic
  model fit (left panel) and an exponential profile fit (right panel).
  In each panel, the dashed curve is the point-source component, the
  dotted curve is the host galaxy model, and the sold curve is the
  combined point source + host galaxy model.  The point-source and
  host galaxy radial profiles were measured from the 2-dimensional
  output files produced by GALFIT using the IRAF ``ellipse'' task.
\label{radplotV}}
\end{figure}

While the exponential disk models were clearly inadequate, the
S\'ersic model fits were much more successful.  The S\'ersic fits
still left substantial residuals within the inner $r \approx
1\arcsec$, but the residuals appeared to be more random, and confined
to smaller radii, than the systematic residuals of the exponential
fits.  After subtracting the best-fitting S\'ersic model, there is no
indication of any residual spiral structure in the host galaxy.  The
magnitudes of the point source and host galaxy, and the S\'ersic
indices and effective radii, from the best-fitting models are given in
Table \ref{galfitresults}.  Combining the measurements from all four
bands, the model fits give a S\'ersic index of $n = 3.6 \pm 0.2$,
close to a de Vaucouleurs law \citep{dev48}, which corresponds to
$n=4$, and $r_e = 590 \pm 50$ pc.  The model fits also gave the
following results (averaged over all four bands): major axis P.A. =
$45\arcdeg \pm 6\arcdeg$, axis ratio $b/a = 0.79 \pm 0.03$, and
boxiness/diskiness parameter = $0.07 \pm 0.10$ (with positive numbers
representing boxy isophotes).  Figure \ref{radplotV} shows the radial
profiles of the galaxy and the exponential and S\'ersic models for the
\emph{V} band; the one-dimensional profiles show that the S\'ersic
model is a better fit than the exponential model for $r \gtrsim
1\arcsec$.

The uncertainties in the derived magnitudes are dominated by
systematic errors in the fits, mainly due to the PSF models and the
possible systematic deviation of the host galaxy from a pure S\'ersic
profile.  We estimate these uncertainties to be of order $\sim0.2$ mag
for the point source and host galaxy magnitudes listed in Table
\ref{galfitresults}.

The dereddened AGN point source colors are consistent with a power-law
slope in the range $\alpha \sim 1 - 1.5$ for $f_\lambda \propto
\lambda^{-\alpha}$, although a single power-law model cannot
simultaneously fit the \emph{BVRI} measurements.

\section{Discussion}

\subsection{Black Hole Mass}

The mass of the black hole in POX 52 can only be estimated using
indirect techniques.  For Seyfert 1 galaxies and quasars, there is a
well-established correlation between the broad-line region radius
(\rblr) and the nonstellar continuum luminosity in the optical, which
has been calibrated by reverberation mapping measurements of the time
delay between continuum and broad-line flux variations \citep{kas00}.
If the BLR gas kinematics are dominated by gravitational motion
(rather than by winds or outflows or other motions), then the
broad-line widths can be combined with the BLR radius to give a virial
estimate of \mbh.  We use the calibration of the BLR size-luminosity
relationship from \citet{kas00}:

\begin{equation}
\rblr = (32.9 \pm 2.0) \left[ \frac{\lambda L_\lambda (5100~
    \mbox{\AA})}{10^{44} \mathrm{~erg~s\per}} \right]^{(0.700 \pm 0.033)}
\mathrm{lt-days},
\label{equation1}
\end{equation}
where $L_\lambda$(5100 \AA) is the luminosity of the nonstellar
continuum at 5100 \AA.
The central mass is then derived from \rblr\ and the \hbeta\ linewidth
according to
\begin{equation}
\mbh = 1.5 \times10^5 \left( \frac{\rblr}{\mathrm{lt-days}} \right)
\left( \frac{\vfwhm}{10^3 \mathrm{~km~s^{-1}}} \right)^2 \msun,
\label{equation2}
\end{equation}
where \vfwhm\ is the FWHM, in velocity units, of the broad component
of \hbeta.  The calibration of the size-luminosity relationship is
subject to some uncertainty, as discussed in detail by \citet{ves02},
as its slope for Seyfert galaxies is somewhat different from that for
quasars, and also dependent on whether or not the discrepant galaxy
NGC 4051 is included in the fitting.  Another issue is how \vfwhm\ is
to be measured.  \citet{kas00} measured FWHM(\hbeta) in the RMS
spectrum from an entire reverberation campaign, which isolates the
variable portion of the emission line.  For POX 52 we have only a
single-epoch measurement.  As shown by \citet{ves02}, single-epoch
measurements of \vfwhm\ for a given AGN tend to cluster around the
value of FWHM(\hbeta, RMS) with roughly $\sim20-25\%$ scatter, so the
use of a single-epoch FWHM should be a reasonable approximation to the
RMS linewidth.

\citet{ves02} has argued that single-epoch measurements of \vfwhm\ for
quasars should be performed on the total \hbeta\ profile, including
any narrow component that might be present. However, POX 52 and NGC
4395 are extreme cases in the sense of having very weak broad
components with strong narrow components that dominate the total
\hbeta\ flux.  For these objects, the FWHM of the \emph{total} \hbeta\
profile is roughly an order of magnitude smaller than that of the
broad component alone, so the width of the full \hbeta\ profile would
have almost no sensitivity to the broad component.  In the Seyferts
and quasars used to determine the linewidth-luminosity correlation,
\hbeta\ is dominated by the broad component, sometimes containing a
weak bump of narrow-line emission.  For an appropriate comparison
between POX 52 and the calibration sample, the best option is to use
only the broad-line component to determine \vfwhm\ and hence \mbh,
following the approach used by \citet{fh03} for NGC 4395.

From the GALFIT decomposition, the extinction-corrected $B$-band
absolute magnitude for the point source component is $M_B = -16.7$
mag, corresponding to $\lambda L_\lambda$(5100 \AA) $=
1.6\times10^{42}$ erg s\per.  Then, for $\vfwhm = 760$ \kms, equations
1 and 2 yield $\rblr \approx 1.8$ lt-days and $\mbh \approx 1.6 \times
10^5$ \msun.  This mass estimate is admittedly quite uncertain.  The
linewidth-mass-luminosity relations have only been calibrated for
masses greater than $10^6$ \msun\ \citep{kas00}, so an
order-of-magnitude extrapolation is required if this method is to be
applied to POX 52.  In addition, there appears to be significant
intrinsic scatter in these relations, and the \hbeta\ method may yield
masses that are uncertain by a factor of $\sim3$
\citep[e.g.,][]{pet03}.  A further caveat is that the imaging and
spectroscopic observations were not simultaneous, and the active
nucleus in POX 52 could be variable on short timescales.
Nevertheless, even bearing these uncertainties in mind, the resulting
mass estimate indicates that the black hole in POX 52 most likely has
a mass well below $10^6$ \msun.

With this estimate, it is possible to examine whether POX 52 follows
the correlations between black hole mass and host galaxy properties.
For the \msigma\ relation, we use the recent fit by \citet{tre02}:
\begin{equation}
\log(\mbh / \msun) = \alpha + \beta \log(\sigmastar / 200~
  \mathrm{km~s\per}), 
\end{equation}
where $\alpha = 8.13 \pm 0.06$ and $\beta = 4.02 \pm 0.32$.  The
galaxies used to calibrate this relation have \sigmastar\ ranging from 67
to 385 \kms.  Extrapolating this correlation to $36 \pm 5$ \kms, the
expected black hole mass is $(1.4 \pm 1.1)\times10^5$ \msun, where the
uncertainty includes both the measurement error on \sigmastar\ for POX 52
and the uncertainty in the fitted parameters of the \msigma\ relation.
The agreement between the black hole mass estimated from FWHM(\hbeta)
and the mass expected from the \msigma\ relation is surprisingly good,
given the substantial uncertainties in both estimates.  With two
independent methods both suggesting a black hole mass of order $10^5$
\msun, POX 52 becomes the second example (after NGC 4395) of a
``dwarf'' AGN with an intermediate-mass black hole.  

The correlation between black hole mass and host galaxy bulge
luminosity (\lbul) can also be compared with this \mbh\ estimate,
since the photometric decomposition suggests that the host galaxy
light is predominantly from a bulge or spheroidal component with
little or no disk contribution.  Using the fit to the \mbh-\lbul\
correlations from \citet{kg01} and from \citet{mh03}, which use
$B$-band bulge luminosities, the implied value of \mbh\ is in the
range $(5.0-7.3)\times10^6$ \msun.  This is more than an order of
magnitude larger than the mass derived from the \hbeta\ linewidth, and
it is an extreme case in terms of the large disagreement between the
\msigma\ relation and the \mbh-\lbul\ relation.  The \msigma\
correlation is generally thought to have smaller scatter than the
\mbh-\lbul\ correlation \citep{fm00, geb00}, although \citet{mh03}
claim that \mbh\ correlates most tightly with bulge mass rather than
with \sigmastar\ or luminosity alone.  There appears to be an extremely
large amount of scatter in the $B$-band \mbh-\lbul\ relation for
$M_\mathrm{B,bul} > -18$ mag; as \citet{kg01} point out, the total
range in \mbh\ is two orders of magnitude for a given value of \lbul.
All of the galaxies currently used to calibrate the \msigma\
correlation have $\sigmastar \geq67$ \kms\ and $\mbh > 2\times10^6$ \msun,
so there are no galaxies with known \mbh\ that are comparable in
overall properties to POX 52.  The best hope for a local calibrator in
this mass range may be the Local Group dwarf elliptical galaxy NGC
205, which has a central velocity dispersion of 30 \kms\ \citep{bpn91}
and $M_B = -15.4$ mag, and is close enough that a measurement of \mbh\
with the \emph{Hubble Space Telescope} should be feasible.

To determine the Eddington ratio $L/\ledd$, we estimate the bolometric
luminosity (\lbol) of the AGN, using the optical continuum luminosity
combined with a bolometric correction based on either the spectral
energy distribution (SED) of NGC 4395, or that of a normal quasar.  If
we assume that POX 52 has an SED similar to that of NGC 4395, then the
ratio of the bolometric luminosities of the two objects will be
similar to the ratio of optical AGN luminosities.  Following
\citet{fh03}, we assume a distance of 4.2 Mpc to NGC 4395.  Using
either the $B$-band continuum flux of the NGC 4395 nucleus
\citep{fs89} or its narrow \hbeta\ flux \citep{kra99}, we find that
the AGN in POX 52 is approximately 200 times more luminous than the
one in NGC 4395.  NGC 4395 has a bolometric luminosity of
$\sim5\times10^{40}$ erg s\per\ \citep[][rescaled to $D = 4.2$
Mpc]{mor99}; this implies $\lbol \approx 1\times10^{43}$ erg s\per\
for POX 52.  Alternatively, applying the standard quasar $B$-band
bolometric correction from \citet{elv94}, the estimated bolometric
luminosity is $2\times10^{43}$ erg s\per.  Then, for $\mbh = 1.6
\times10^5$ \msun, the Eddington ratio is in the range $\lbol/\ledd
\approx 0.5 - 1$.  Thus, the black hole in POX 52 appears to be
radiating at nearly its maximal rate, and it may be undergoing a major
episode of mass accretion.  In contrast, \citet{fh03} estimated
$\lbol/\ledd \approx 2\times10^{-2} - 2\times10^{-3}$ for NGC 4395.
Given the apparently large difference in the accretion rates of these
two objects, the overall similarity in their optical spectra is
somewhat surprising.  

Assuming that $\lbol \leq \ledd$ in POX 52, a lower limit to \mbh\ can
be derived.  The smaller estimate of \lbol\ ($1\times10^{43}$ erg
s\per) implies $\mbh \geq 7\times10^4$ \msun.

\subsection{Host Galaxy and Environment}

What type of galaxy is POX 52?  The GALFIT profile measurements
indicate that it is unlikely to be a pure disk galaxy.  Its absolute
magnitude and effective radius are within the normal range for dwarf
ellipticals, and its $(B-V)$ color of 0.8 mag is also typical of dwarf
ellipticals \citep{gav01}.  However, the S\'ersic index of $3.6 \pm
0.2$ is outside the usual range of $\sim1.0 - 3.0$ for dwarf
ellipticals in Virgo \citep{ggv03}.

\begin{figure}
\begin{center}
\scalebox{0.45}{\includegraphics{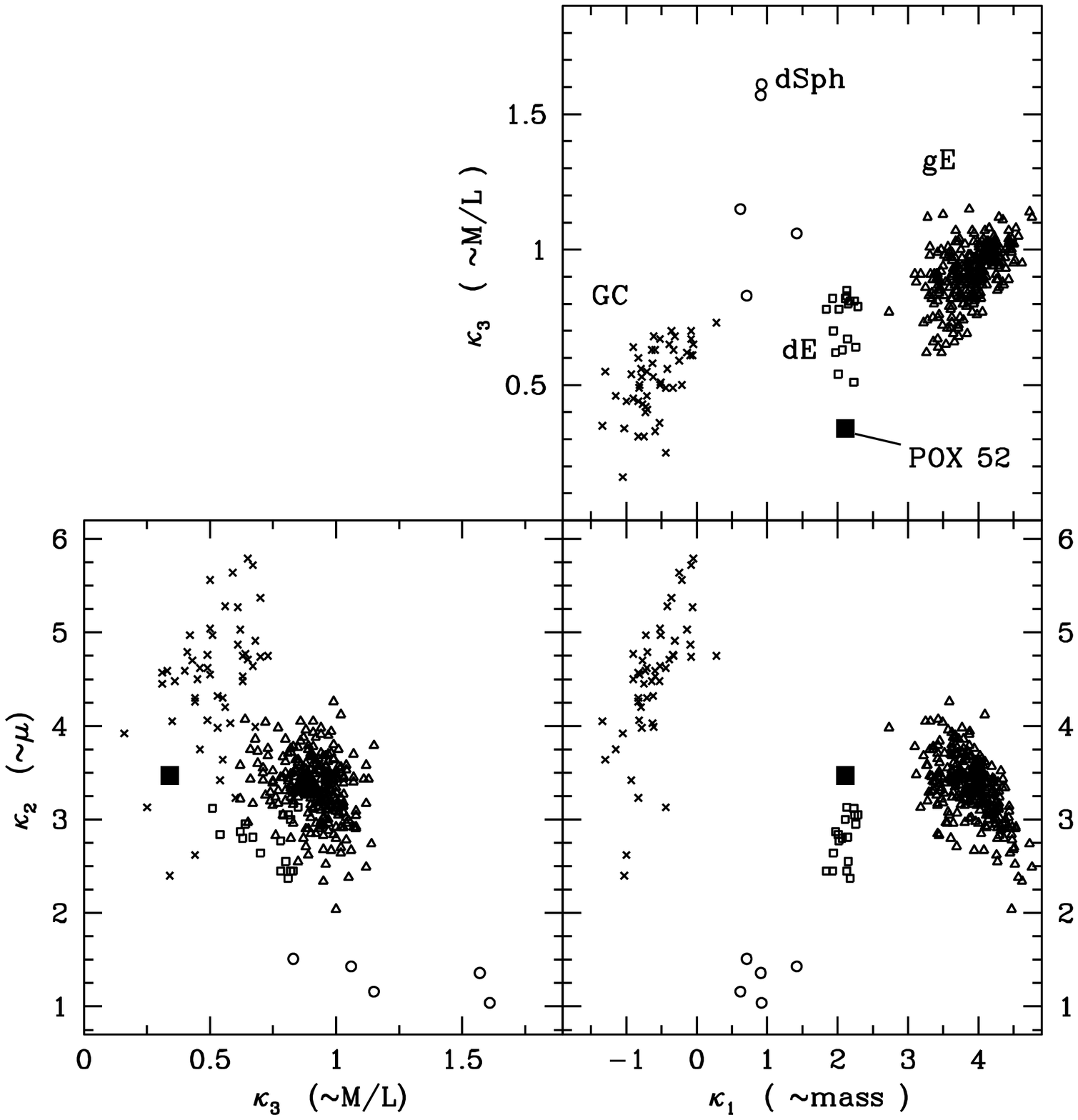}}
\end{center}
\caption{``$\kappa$-space'' projections of the fundamental plane,
  following \citet{bur97} and \citet{ggv03}.  The parameters
  $\kappa_1$, $\kappa_2$, and $\kappa_3$ are related to galaxy mass,
  surface brightness, and mass-to-light ratio.  Data points taken from
  \citet{bur97} include giant elliptical galaxies (triangles), dwarf
  spheroidals (circles), and globular clusters (crosses).  Dwarf
  elliptical galaxies (open squares) are from \citet{ggv03}.  POX 52
  is shown as a large filled square.
\label{kappa}}
\end{figure}

To examine whether POX 52 is closer in properties to dwarf or giant
ellipticals, it is useful to determine its location in the fundamental
plane.  We use the ``$\kappa$-space'' projections of the fundamental
plane, as defined by \citet{bbf92} and \citet{bur97}. In this
projection, the fundamental plane is viewed along axes that correspond
to galaxy mass ($\kappa_1$), surface brightness ($\kappa_2$), and
mass-to-light ratio ($\kappa_3$).  Dwarf ellipticals, giant
ellipticals, dwarf spheroidals, and globular clusters are
well-segregated in $\kappa$-space, as shown by \citet{bur97} and
\citet{ggv03}.  Combining our $B$-band host galaxy parameters with the
stellar velocity dispersion, we find $\kappa_1 = 2.11$, $\kappa_2 =
3.47$, and $\kappa_3 = 0.34$ for POX 52.  Figure \ref{kappa}
illustrates the position of POX 52 in the $\kappa$-space projections,
in comparison with data from the literature.  POX 52 lies closest to
the family of dwarf ellipticals, although it is slightly offset from
the Virgo dwarf ellipticals studied by \citet{ggv03} in the sense of
having somewhat higher surface brightness and lower mass-to-light
ratio.  One possible explanation could be a younger stellar
population in POX 52 than in the Virgo dwarf ellipticals, but this may
be unlikely given the normal $(B-V)$ color of POX 52.  A more likely
explanation may be systematic uncertainty in the GALFIT decomposition;
the unusual S\'ersic index for POX 52 could be the result of an
imperfect model fit to the host galaxy resulting from the presence of
the extremely bright central point source.  Imaging data from the
\emph{Hubble Space Telescope} would provide a much more accurate
decomposition into point-source and host galaxy components and a
better measurement of the S\'ersic index and the galaxy profile within
the inner arcsecond.  With the present data, we conclude that POX 52
is best classified as a dwarf elliptical, although perhaps a peculiar
one.

To our knowledge, no other dwarf elliptical galaxy has been found to
host a Seyfert nucleus.  One object that has been referred to as a
dwarf elliptical with an AGN is the nearby galaxy NGC 3226
\citep{geo01}.  Its nucleus is a LINER with a broad component to the
\hal\ emission line \citep{hfs97}, and it is also an X-ray source
\citep{geo01}.  While there is no question that NGC 3226 hosts a
genuine AGN, it is too massive and too large to be classified as a
true dwarf elliptical galaxy.  It has an absolute magnitude of $M_B =
-19.4$ and $\sigmastar \approx 180$ \kms\ \citep{sp98}, which suggests a
likely black hole mass of order $10^8$ \msun\ from the \msigma\
relation.  Thus, NGC 3226 appears to be a small but ordinary
elliptical, probably with a fairly massive black hole that is
accreting at an extremely sub-Eddington rate, and it is very different
from the ``dwarf AGNs'' in POX 52 and NGC 4395 with $\mbh < 10^6$
\msun.

\citet{nw96} found a correlation between the stellar and gaseous
linewidths in Seyfert galaxies.  The near equality between
FWHM([\ion{O}{3}]) and $2.35\sigmastar$ in most Seyferts suggests that
the widths of the narrow lines are largely determined by virial motion
in the bulge of the host galaxy, although objects with radio jets tend
to have systematically broader emission lines for a given \sigmastar.
Interestingly, POX 52 follows this trend almost perfectly, with
$2.35\sigmastar = 85 \pm 12$ \kms, nearly equal to the [\ion{O}{3}]
FWHM of $87 \pm 10$ \kms.  This result extends the Nelson \& Whittle
correlation to Seyfert galaxies with velocity dispersions below 40
\kms, which may be of use for applying the \msigma\ relation to other
POX 52-like objects whose stellar continua are too faint or too
diluted by AGN emission for \sigmastar\ to be measured directly.  (KSB
reached a different conclusion regarding the NLR dynamics, due to
their erroneous measurement of the emission-line widths.)

We note that POX 52 is probably not a Wolf-Rayet galaxy, contrary to
the conjecture by KSB.  In Wolf-Rayet galaxies, the emission bump
surrounding 4686 \AA\ is a blend of several features including
\ion{N}{3} $\lambda4640$, [\ion{Fe}{3}] $\lambda4658$, [\ion{Ar}{4}]
$\lambda4711$, and [\ion{Ar}{4}] $\lambda4740$, in addition to
\ion{He}{2} \citep[e.g.,][]{ks81,sck99}.  None of these features is
present in POX 52.  Also, the Wolf-Rayet emission bump from \ion{C}{4}
$\lambda5808$ does not appear in POX 52.  There is no other evidence
for recent star formation either from spectral features or from the
host galaxy colors or morphology.   The broad component to \ion{He}{2}
$\lambda4686$ most likely has its origin in the broad-line region of
the AGN, rather than in Wolf-Rayet stars.

POX 52 is not a member of a cluster, but a search of the NED database
reveals that it lies close to a known galaxy group.  The group USGC
S178, with $\langle cz \rangle = 6587$ \kms, is centered at $\alpha =
12^\mathrm{h} 00^\mathrm{m} 10\fs1$, $\delta = -20\arcdeg 27\arcmin
58\arcsec$ \citep{ram02}.  This group has a velocity dispersion of 243
\kms\ for the 5 galaxies that are known to be members.  POX 52 is not
listed by Ramella \etal\ as a group member, but its recession velocity
fits within this velocity range, and its projected distance from the
group center is 1.3 Mpc, so it is may be associated with the group
although it is well outside the group's central region.  The local
environment does not, however, yield any clues as to why this
particular dwarf galaxy should be undergoing an AGN accretion episode.

\section{Conclusions}

New spectra confirm that POX 52 is a genuine Seyfert galaxy, and the
broad components seen on the permitted line profiles (first detected
by KSB) demonstrate that POX 52 should be classified as a Seyfert 1.
The host galaxy morphology is most similar to that of a dwarf
elliptical.  This is only the second known example of a Seyfert
nucleus in a dwarf galaxy, and the first in a dwarf elliptical.  The
black hole mass, estimated from the \hbeta\ linewidth-luminosity-mass
correlations for Seyferts, is $1.6\times10^5$ \msun, and for this
value of \mbh\ POX 52 falls very close to the extrapolated \msigma\
relation of nearby galaxies.  If this black hole mass estimate is
accurate, then the AGN in POX 52 is likely to be radiating at nearly
its maximal rate, with $\lbol\sim(0.5-1)\times\ledd$.

With only two known AGNs in dwarf galaxies, little can be said about
the statistics or demographics of this class of objects.  The small
sample can be at least partly attributed to the fact that no
systematic searches for AGNs in dwarf galaxies have been carried out.
Additional examples could potentially be detected in objective-prism
surveys for emission-line galaxies such as the KISS survey
\citep{sal00}.  Currently, the Sloan Digital Sky Survey presents
perhaps the best opportunity to find more members of this apparently
rare class of AGNs.  A search of the Sloan data archives to select
additional Seyfert nuclei in dwarf galaxy hosts is currently underway
\citep{gh04}, as a first step toward filling out the census of massive
black holes with $M < 10^6$ \msun.

\acknowledgments 

We thank George Becker for assisting with the Keck observations, and
Marla Geha and Gary Ferland for helpful discussions.  Research by
A.J.B. is supported by NASA through Hubble Fellowship grant
\#HST-HF-01134.01-A awarded by STScI, which is operated by AURA, Inc.,
for NASA, under contract NAS 5-26555.  W.L.W.S. acknowledges support
from NSF grant AST-0206067.  This research has made use of the
NASA/IPAC Extragalactic Database (NED) which is operated by the Jet
Propulsion Laboratory, California Institute of Technology, under
contract with the National Aeronautics and Space Administration.  This
research has also made use of data obtained from the High Energy
Astrophysics Science Archive Research Center (HEASARC), provided by
NASA's Goddard Space Flight Center.  Some data presented herein were
obtained at the W.M. Keck Observatory, which is operated as a
scientific partnership among Caltech, the University of California,
and NASA. The Observatory was made possible by the generous financial
support of the W.M. Keck Foundation.  The authors wish to recognize
and acknowledge the very significant cultural role and reverence that
the summit of Mauna Kea has always had within the indigenous Hawaiian
community.  We are most fortunate to have the opportunity to conduct
observations from this mountain.

\end{document}